\documentclass[conference]{IEEEtran}
\IEEEoverridecommandlockouts

\usepackage{tabularx}
\usepackage{tikz}
\usepackage{pgfplots}
\usepackage{caption}

\pgfplotsset{compat=newest} 
\usepackage{algorithm}
\usepackage{algorithmic}
\usepackage{graphicx}
\usepackage{diagbox}
\usepackage{adjustbox}
\usepackage{balance}
\usepackage{csquotes}
\usepackage{framed}
\usepackage{multirow}
\usepackage{url}
\usepackage{amsmath}
\usepackage[table,xcdraw]{xcolor}
\usepackage[many,listings]{tcolorbox}
\tcbuselibrary{breakable}
\usepackage{listings}
\usepackage{xcolor} 

\usepackage{booktabs}
\usepackage{hyperref}
\usepackage{tcolorbox}
\usepackage[utf8]{inputenc}
\pgfplotsset{compat=1.17}
\definecolor{myblue}{rgb}{0.4, 0.6, 0.8}
\definecolor{myyellow}{rgb}{1.0, 0.9, 0.6}
\definecolor{myorange}{rgb}{1.0, 0.7, 0.5}
\definecolor{mygreen}{rgb}{0.6, 0.8, 0.6}
\definecolor{tablegray}{HTML}{F6F6FA}

\definecolor{bg}{rgb}{1,1,1}
\definecolor{keyword}{rgb}{0.75,0,0}
\definecolor{string}{rgb}{0.58,0,0.82}
\definecolor{redtext}{rgb}{0.75,0,0}
\definecolor{numberbg}{rgb}{0.8,0.8,0.84}
\usepackage[justification=raggedright, singlelinecheck=false]{caption}
\usepackage[normalem]{ulem}
\usepackage{xcolor}

\definecolor{myteal}{RGB}{0, 128, 128} 
\tcbuselibrary{skins}
\tcbset{RQsboxstyle/.style={
    width=\columnwidth,       
    colframe=myteal,            
    colback=myteal!10,          
    coltitle=black,           
    boxrule=0.1mm,            
    arc=1mm,                  
    auto outer arc,           
    boxsep=1mm,               
    drop shadow={black!50!white},  
    leftrule=1.5mm
}}

\tcbset{blueboxstyle/.style={
    width=\columnwidth,       
    colframe=blue!50!black,        
    colback=blue!5,        
    coltitle=black,          
    boxrule=0.1mm,           
    arc=1mm,                  
    auto outer arc,           
    boxsep=1mm,               
    drop shadow={black!50!white} 
}}

\tcbset{yellowboxstyle/.style={
    width=\columnwidth,       
    colframe=yellow!40!black,          
    colback=yellow!10,        
    coltitle=black,           
    boxrule=0.1mm,            
    arc=1mm,                  
    auto outer arc,           
    boxsep=1mm,               
    drop shadow={black!50!white}, 
    leftrule=1.5mm
}}

\usepackage{tikz}
\usetikzlibrary{backgrounds}
\usepackage{graphicx}

\definecolor{myexact}{HTML}{C3B1E1}  
\definecolor{mynear}{HTML}{FFB347}   
\definecolor{myunique}{HTML}{B2D8B2} 

\definecolor{trainbg}{gray}{0.95}    
\definecolor{testbg}{gray}{0.90}     

\begin{document}



\title{Evaluating Large Language Models for Security Bug Report Prediction}

\author{
    \IEEEauthorblockN{Farnaz Soltaniani}
    \IEEEauthorblockA{Technische Universität Clausthal\\
    Germany\\
    farnaz.soltaniani@tu-clausthal.de}
    \and
    \IEEEauthorblockN{Shoaib Razzaq}
    \IEEEauthorblockA{Technische Universität Clausthal\\
    Germany\\
    shoaib.razzaq@tu-clausthal.de}
    \and
    \IEEEauthorblockN{Mohammad Ghafari}
    \IEEEauthorblockA{Technische Universität Clausthal\\
    Germany\\
    mohammad.ghafari@tu-clausthal.de}
}

\maketitle

\begin{abstract}
Early detection of security bug reports (SBRs) is critical for timely vulnerability mitigation.
We present an evaluation of prompt-based engineering and fine-tuning approaches for predicting SBRs using Large Language Models (LLMs).
Our findings reveal a distinct trade-off between the two approaches. 
Prompted proprietary models demonstrate the highest sensitivity to SBRs, achieving a G-measure of 77\% and a recall of 74\% on average across all the datasets, albeit at the cost of a higher false-positive rate, resulting in an average precision of only 22\%.
Fine-tuned models, by contrast, exhibit the opposite behavior, attaining a lower overall G-measure of 51\% but substantially higher precision of 75\% at the cost of reduced recall of 36\%.
Though a one-time investment in building fine-tuned models is necessary, the inference on the largest dataset is up to 50 times faster than that of proprietary models.
These findings suggest that further investigations to harness the power of LLMs for SBR prediction are necessary.

 

%

\end{abstract}


\section{INTRODUCTION}\label{sec:introduction}

Bug reports (BRs) are the primary means to inform development teams of a problem or defect found in software. 
Issue reports with security implications, known as security bug reports (SBRs), require higher fix priority.
However, identifying these reports is challenging, they often progress slowly, and many remain unresolved for extended periods~\cite{Noah2022}. 
Furthermore, limited security knowledge among mainstream developers often leads to SBRs being misclassified as non-security bug reports (NSBRs) during reporting~\cite{5463340}.



Researchers have adopted AI to classify SBRs from NSBRs, primarily based on the description field of bug reports~\cite{soltaniani2025security, FARSEC, 9371393, Ohiraetal, DeeplearningBasedSoftwareBugClassification2024, 10.1145/3643991.3644903}. 
However, the scarcity of SBRs in large and diverse datasets leads to severe class imbalance, which limits the effectiveness of predictive models. In previous work~\cite{soltaniani2025security}, we conducted a comparative study of BERT and state-of-the-art models for SBR prediction. Extensive experiments across multiple scenarios, including within-project and cross-project settings, demonstrated the superiority of BERT when sufficient training data were available.

Building on this prior work and motivated by the growing attention to Large Language Models (LLMs), this paper investigates the potential of both proprietary pre-trained models and open-source, small fine-tuned LLMs for SBR prediction. In particular, we aim to address the following two research questions:


\textbf{RQ\textsubscript{1}:} How effective are proprietary LLMs at predicting SBRs?

We assess how effectively proprietary LLMs, specifically GPT \cite{achiam2023gpt} and Gemini \cite{geminiteam2025}, can predict SBRs.
We provide a prompt that defines the task of SBR and NSBR identification, includes the description of a bug report, and specifies the desired output format.
The evaluation of these models revealed a distinct divergence in classification strategies between GPT and Gemini. Gemini consistently prioritized sensitivity, acting as a robust detector with an average recall of 0.74 across all tested datasets, identifying more than three times as many SBRs as GPT. However, this high sensitivity came at the cost of a high false positive rate (FPR), resulting in a lower precision of 0.22.
Conversely, GPT adopted a highly conservative approach, failing to flag the majority of SBRs, with a recall of 0.23 and a precision of 0.26.

\begin{table*}[htb]
\footnotesize
\centering
\setlength{\tabcolsep}{18pt}
\renewcommand{\arraystretch}{1.3}
\caption{Dataset-Related Information, Where SBRs Indicate the Proportion of Security Bug Reports.}
\scalebox{1.0}{
\begin{tabular}{|l | l | c | c | c|}
\hline
\textbf{Project} & \textbf{Description} & \textbf{BRs} & \textbf{NSBRs} & \textbf{SBRs}\\
\hline\hline
Chromium & Web browser called Chrome & 41,940 & 41,132 & 808 \\ \hline
Derby & A relational database management system & 1,000 & 821 & 179 \\ \hline
Camel & A rule-based routing and mediation engine & 1,000 & 926 & 74 \\ \hline
Ambari & Hadoop management web UI backed by its RESTful APIs & 1,000 & 944 & 56 \\ \hline
Wicket & Component-based web application framework for Java programming & 1,000 & 953 & 47 \\ \hline
\end{tabular}}
\label{tab:datasetsDetails}
\end{table*}


\textbf{RQ\textsubscript{2}:} How effective are fine-tuned LLMs at predicting SBRs?

We investigate the effectiveness of two fine-tuned small LLMs for SBR prediction: (1) encoder-only LLMs, such as BERT \cite{devlin2019bert} and its variant DistilBERT \cite{distilbert}; (2) decoder-only LLMs, such as DistilGPT \cite{su2024distilledgptsourcecode} and Qwen \cite{qwen2.5}, which we adapt for prediction using the final token representation.
%
%
DistilBERT consistently achieved the highest prediction performance, notably surpassing BERT with a G-measure of 0.51. In general, however, the effectiveness of all models varied substantially across datasets. Performance was optimal on larger datasets and those with higher SBR-to-NSBR ratios, but declined on smaller or more imbalanced datasets.

In summary, we show that the best proprietary LLM achieves 26\% and 38\% higher G-measure and recall, respectively, than those of the best fine-tuned model, but at the cost of a higher false positives. In contrast, the best fine-tuned model attains 53\% higher precision and offers substantially lower inference latency.
These findings highlight the need for further in-depth studies such as identifying the factors that contribute to successful SBR classification by the models.


The remainder of this paper is structured as follows.
Section \ref{sec:RelatedWork} presents the related work. Section \ref{sec:Methodology} describes the methodology. Section \ref{sec:EXPERIMENTRESULTS} presents the results, and Section \ref{sec:Discussion} discusses our findings. Finally, Section \ref{sec:ThreatsToValidity} lists the threats to validity of this study and Section \ref{sec:Conclusion} concludes the paper.

\section{Related Work}
\label{sec:RelatedWork}

Researchers have proposed text-based machine learning (ML) approaches to improve the classification of BRs as either SBRs or NSBRs in bug-tracking systems \cite{Zaher2021, 6473768, soltaniani2025security}. However, the scarcity of SBRs leads to a significant class imbalance, resulting in sparse feature representations that degrade the performance of traditional ML models.
To address this issue, researchers suggest preprocessing datasets before applying ML algorithms \cite{FARSEC, shu2019better, LTRWES}. This preprocessing may involve filtering out NSBRs that contain security-related keywords \cite{FARSEC}, using content-based filtering to minimize the risk of misclassifying NSBRs as SBRs \cite{LTRWES}, utilizing the Synthetic Minority Oversampling Technique (SMOTE) and its variant, SMOTUNED \cite{shu2019better}, or applying a k-means clustering approach. The k-means method maintains diversity among NSBRs by grouping data based on structure and selecting the samples nearest to each cluster's centroid \cite{CASMS}.

To exemplify, Peters et al. \cite{FARSEC} introduced the FARSEC framework, designed to improve SBR prediction by filtering out misleading NSBRs. The framework identifies the top 100 terms relevant to SBRs and scores NSBRs based on the frequency and significance of these terms. NSBRs with scores above a specific threshold (e.g., 0.75) are filtered out as likely misclassifications. Subsequently, five ML algorithms were used to predict the probability of a bug report being security-related.
Wu et al. \cite{9371393} built upon the work of Peters et al. \cite{FARSEC} to improve label accuracy across the same five datasets. They demonstrated that using the same prediction models as Peters et al., simple text classification methods outperformed both the hyperparameter-tuned models and the data preprocessing techniques applied by Shu et al. \cite{shu2019better}. Notably, Random Forest (RF) exhibited the best overall performance, highlighting its effectiveness for this task.

Soltaniani et al. \cite{soltaniani2025security} conducted a comprehensive comparative study of BERT and RF for SBR prediction across within-project and cross-project scenarios. While RF outperformed BERT in standard within-project evaluations, BERT proved superior when training data was augmented with diverse bug reports, achieving a G-measure of 66\%. Furthermore, BERT demonstrated stronger generalization in cross-project prediction, attaining a G-measure of 62\%. The study also employed the FARSEC keyword-filtering approach, finding that it had a limited or mixed impact on the models' performance.

Despite the growing body of literature on the usage of different AI models in predicting SBRs, the majority of existing work focuses on traditional supervised ML approaches, with a few studies examining BERT \cite{soltaniani2025security}. As a result, the effectiveness of LLMs using either fine-tuning or prompting and their relative strengths remain largely underexplored.
We evaluate the effectiveness of fine-tuned and proprietary LLMs for the task of SBR prediction.

\section{Methodology}
\label{sec:Methodology}

In this section, we first introduce the datasets, then describe our train-test split methodology, present the model architectures employed, and detail the evaluation metrics used in this study.

\subsection{Dataset}\label{sec:DatasetSelection}

Previous studies on SBR prediction \cite{Ohiraetal, FARSEC, 9371393, LTRWES, CASMS} have extensively relied on five datasets, namely Chromium, Ambari, Wicket, Derby, and Camel, for security bug prediction. We use the refined version of these datasets \cite{9371393}, listed in Table \ref{tab:datasetsDetails}. Each row corresponds to a bug report, with columns showing its features. We collected the essential features of bug ID, description, summary, and security. We combined the description and summary features to consolidate information into a single text field.

\begin{figure}[htbp]
    \centering
    \vspace{4pt} 
    \includegraphics[width=0.4\textwidth]{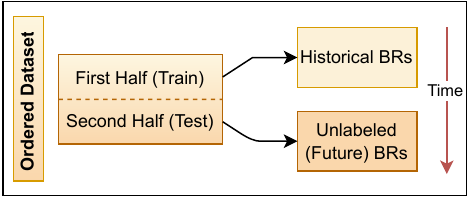}
    \centering
    \caption{Sequential Dataset Partitioning.}
    \label{fig:historicalBugReports}
\end{figure}

\subsection{Train and Test Splits}\label{sec:Evaluation}

One fundamental assumption of traditional ML approaches is that the training (past) data and present (test) data have the same features and distribution. In real-world settings, however, this assumption rarely holds. Accordingly, we design our experimental setup to capture this temporal mismatch. As illustrated in Figure \ref{fig:historicalBugReports}, we sort each dataset chronologically and split it into two equal parts. The first half contains historical bug reports used for training, while the second half comprises future, unlabeled reports used for evaluation.

The distribution of SBRs and NSBRs for the chronologically ordered datasets is shown in Table \ref{tab:TreatmentsAndDatasets-WPP}. 
During the training process of fine-tuned models, the training subset is further divided into 10\%  validation and 90\% training for parameter tuning. After completing the training, we assess the model's performance on the testing subset.


\begin{table}[htpb]
\caption{Distribution of SBRs and NSBRs.}
\label{tab:TreatmentsAndDatasets-WPP}
\footnotesize
\centering
\setlength{\tabcolsep}{8pt}
\renewcommand{\arraystretch}{1.3}
\begin{tabular}{|l|cc|cc|}
\hline
\multirow{2}{*}{\textbf{Target Dataset}} & \multicolumn{2}{c}{\textbf{Train Set}} & \multicolumn{2}{c|}{\textbf{Test Set}} \\ \cline{2-3} \cline{4-5}
 & \textbf{SBR} & \textbf{NSBR} & \textbf{SBR} & \textbf{NSBR} \\ \hline\hline
Chromium & 371 & 20599 & 437 & 20533 \\ \hline
Derby    & 82  & 418   & 97  & 403   \\ \hline
Camel    & 28  & 472   & 46  & 454   \\ \hline
Ambari   & 40  & 460   & 16  & 484   \\ \hline
Wicket   & 24  & 476   & 23  & 477   \\ \hline
\end{tabular}
\end{table}



\subsection{Proprietary LLMs}

We employ Gemini \cite{geminiteam2025} and GPT \cite{achiam2023gpt}, which have both consistently demonstrated strong performance on public benchmarks (e.g., HumanEval \cite{HumanEval}, MBPP \cite{yu2024humaneval}).
GPT is selected for its robust reasoning capabilities, while Gemini is chosen for its substantial free-tier credits (\$300 of initial credits).
These models and their respective versions are listed in Table \ref{tab:LLMVersions}. We access these models via API, which facilitates seamless integration with various applications.\\

\begin{table}[htbp]
    \centering
    \renewcommand{\arraystretch}{1.3}
    \caption{Overview of selected LLMs.}
    \begin{tabular}{|l|l|}
    \hline 
    \textbf{Prompted LLMs} & \textbf{Version} \\ \hline \hline
    OpenAI's GPT &  GPT-4.1 (v. 2025-04-14)\\ \hline
    Google's Gemini & gemini-2.5-flash (v. 2025-06-17)\\ \hline
    \end{tabular}
    \label{tab:LLMVersions}
\end{table}

OpenAI's documentation \cite{openai2023cheatsheet} recommends a `temperature' of 0.2 and `top\_p' of 0.1 for more deterministic and focused results. Accordingly, we used these values as the default settings for both models, as applying identical parameters ensures a controlled comparison.
In this analysis, we evaluate the prompted models solely on the test datasets, both to maintain comparability with the fine-tuned models and to avoid any potential data leakage from the training sets.

\emph{Prompt Development.}
While tailored prompt engineering for each specific LLM may improve individual performance. To mitigate the risk of prompt-induced bias and ensure a controlled evaluation environment, we developed a standardized task-based prompt. Specifically, our prompt encompasses the detection workflow into a single, cohesive instruction set that leverages two key techniques:

\emph{Domain-Specific Role Playing:} Initially, the models are prompted to assume the role of a senior security auditor, as role-playing enhances the accuracy and relevance of the responses by
focusing on a specific domain \cite{kong2024betterzeroshotreasoningroleplay}. 

\emph{Rule-based Verification:} We define a strict step-by-step task and ask the models to independently determine, within a given bug report, whether the report is a security bug SBR or NSBR.
Furthermore, we enforce a rigid, structured output format, preventing the model from generating verbose or ambiguous explanations.
Particularly for each sample in each dataset, we query the models with the prompt in listing~\ref{lst:prompt}.\\

\definecolor{promptteal}{HTML}{008B8B} 
\definecolor{promptbg}{HTML}{F0F2F5} 

\lstdefinestyle{promptstyle}{
    basicstyle=\small\rmfamily,
    columns=fullflexible,   
    keepspaces=true,
    breaklines=true,
    breakatwhitespace=true,
    showstringspaces=false,
    breakindent=0pt,
    escapechar=!,
    frame=none,                
    backgroundcolor=\color{promptbg},
    moredelim=**[is][\bfseries]{(@}{@)},
    moredelim=**[is][\color{blue}\itshape]{(*}{*)}
}

\newtcolorbox{promptbox}{
    enhanced,
    colback=promptbg,       
    colframe=promptteal,    
    boxrule=0pt,           
    leftrule=4pt,          
    arc=0pt,              
    outer arc=0pt,
    left=5pt, right=5pt, top=5pt, bottom=5pt,
    parbox=false,
    nobeforeafter,
    grow to left by=0.3cm,
}
\vspace{0.5em}
\label{lst:prompt}
\begin{promptbox}
\begin{lstlisting}[style=promptstyle]
(@- Role:@) You are a Security Vulnerability Classification Expert (CWE/CVE/STRIDE/MITRE ATT&CK).

(@- Definitions:@)
1) Security: The bug describes a vulnerability (e.g., null pointer, improper memory handling, privacy leakage, improper permission, etc.).

2) Non-security: The bug is a functional error, a usability issue, a UI glitch, a performance issue, or a feature request, with no clear exploit path.

(@- Task:@) determine if the bug report is security-relevant.

(@- Output only this JSON (no extra text):@)
{
  ``explanation'': ``concise reason (<= 20 words)'',
  ``classification'': ``security'' | ``non-security''
}

(@- BUG REPORT:@)
 {Text}
\end{lstlisting}
\end{promptbox}
\captionof{lstlisting}{The Prompt.}

\subsection{Fine-Tuned LLMs}

We select four models for supervised fine-tuning, including BERT, DistilBERT, DistilGPT-2, and Qwen2.5-Coder. We focus on small-sized LLMs due to their suitability for deployment in resource-constrained and security-preserving applications.
Among the evaluated models, BERT and DistilBERT are encoder-based architectures, while DistilGPT-2 and Qwen2.5B are based on decoder architectures.
To our knowledge, distilled variants such as DistilBERT and DistilGPT-2 have not been systematically studied for SBR prediction, offering the advantages of reduced model size and fewer hyperparameters. Qwen2.5-Coder is included due to its strong performance on code intelligence benchmarks and efficient parameter adaptation via LoRA.
We include BERT as a strong baseline for its robust cross-project performance, previously shown to outperform RF, which lacks generalizability. Similarly, we did not employ oversampling techniques (e.g., SMOTE) or filtering methods such as FARSEC, since prior work found these approaches yield limited improvements in SBR prediction~\cite{soltaniani2025security}.

We utilized the Differential Evolution (DE) algorithm to perform an iterative search for optimal hyperparameters (i.e., learning rate, batch size, sequence length, dropout) across four fine-tuned models. For each run, DE was conducted with a population of 20 over 50 generations. Optimization was terminated if no improvement was observed for 10 generations. 


We consider two BERT-based architectures for SBR prediction: the standard BERT model \cite{devlin2019bert}, designed to capture both the context and inherent structure of source code and its distilled variant, namely DistilBERT \cite{distilbert}.

\subsubsection{BERT} Among encoder-only models, BERT is trained using two key objectives: masked language modeling (MLM) and next-sentence prediction (NSP). In MLM, certain tokens in the input sequence are masked and predicted based on the surrounding context, while NSP determines whether two sentences appear consecutively in the original text. Fine-tuning BERT involves adding a simple classification layer to the pre-trained model, allowing all parameters to be adjusted for specific tasks. For model adaptation, we utilized the implementation methodology employed by Soltaniani et al.~\cite{soltaniani2025security}.
We tokenized inputs using the \textit{BertTokenizer}, producing subword token IDs and attention masks. Padding is applied dynamically within each batch, and attention masks ensure the model attends only to meaningful tokens. These token IDs and masks are wrapped in a TensorDataset and loaded via a DataLoader with random sampling for training and evaluation.

\subsubsection{DistilBERT} We employ DistilBERT (Distilled Bidirectional Encoder Representations from Transformers) \cite{distilbert}, a distilled version of the BERT architecture operating on the principle of knowledge distillation, designed to reduce model size and inference latency while retaining the majority of the original model's performance. 
We adapt the pre-trained \textit{distilbert-base-uncased} model for binary classification. We instantiate the model using the \textit{AutoModelForSequenceClassification} architecture, which appends a linear classification head on top of the encoder's pooled output. 
For input processing, we utilize the corresponding DistilBERT tokenizer and the AdamW optimizer with a weight decay of 0.01 to mitigate overfitting and train for 10 epochs.


\subsubsection{DistilGPT2} We evaluate DistilGPT2, a distilled version of the GPT-2 architecture\cite{radford2019language} which applies knowledge distillation to the GPT-2, resulting in a compact network of approximately 82 million parameters. 
We fine-tune the model using the \textit{GPT2ForSequenceClassification} framework, which places a linear classification head on the final token embedding. Since GPT-2 lacks a native padding token, we designate the End of Sequence token as the padding token to enable effective batch processing.
Inputs are first tokenized to capture extended context. The model is then trained for 10 epochs. 

\subsubsection{Qwen2.5-Coder} 
We evaluate Qwen2.5-Coder-0.5B, a lightweight code-oriented decoder-only model designed for software engineering tasks \cite{qwen2.5}. Although primarily intended for code-related applications, bug reports often contain a mixture of natural language, stack traces, and code snippets; therefore, we include this model to examine whether its code-focused pretraining provides an advantage for this hybrid input format. 
We fine-tune the model using LoRA on the \textit{Qwen2.5-Coder-0.5B-Instruct} backbone, assigning the EOS token as padding due to the lack of a native pad token. Inputs are tokenized using the Qwen \emph{AutoTokenizer}, and training is conducted for 10 epochs with an effective batch size and varying learning rates, selecting the best checkpoint based on validating the G-measure.

\subsection{Performance Metrics}

For each bug report, the prediction result can yield four possible outcomes, as detailed in Table \ref{tab:ConfusionMatrix-table}, from which the performance metrics are derived.

\begin{table}[htpb]
\centering
\caption{Confusion Matrix.}
\label{tab:ConfusionMatrix-table}
\renewcommand{\arraystretch}{1.3}
\resizebox{0.95\columnwidth}{!}{%
\begin{tabular}{lp{0.2\columnwidth}p{0.2\columnwidth}p{0.2\columnwidth}} 
 & \multicolumn{1}{c}{}  & \multicolumn{2}{c}{\textbf{Prediction}}                             \\ \cline{3-4} 
 & \multicolumn{1}{l|}{} & \multicolumn{1}{l|}{\textbf{SBRs}} & \multicolumn{1}{l|}{\textbf{NSBRs}} \\ \cline{2-4} 
\multicolumn{1}{l|}{\multirow{2}{*}{\textbf{Actual}}} & \multicolumn{1}{l|}{\textbf{SBRs}}  & \multicolumn{1}{l|}{True Positive (TP)} & \multicolumn{1}{l|}{False Negative (FN)} \\ \cline{2-4} 
\multicolumn{1}{l|}{}                         & \multicolumn{1}{l|}{\textbf{NSBRs}} & \multicolumn{1}{l|}{False Positive (FP)} & \multicolumn{1}{l|}{True Negative (TN)} \\ \cline{2-4} 
\end{tabular}%
}
\end{table}

The evaluation metrics used in this study are as follows: \textit{Precision} measures the fraction of actual SBRs among the predicted SBRs (Equation \ref{eq:precision}). \textit{Recall} assesses the proportion of correctly classified SBRs among all verified SBRs (Equation \ref{eq:recall}). The \textit{F1-score} is the harmonic mean of \textit{Precision} and \textit{Recall} (Equation \ref{eq:f1score}). The \textit{Probability of False Alarm (FPR)} quantifies the fraction of NSBRs mistakenly identified as SBRs, representing the false positive rate (Equation \ref{eq:pf}). Finally, the \textit{G-measure} combines \textit{Recall} with the complement of FPR, offering a balanced evaluation of a model’s performance by capturing its effectiveness in identifying true positives while minimizing false alarms (Equation \ref{eq:gmeasure}).

\begin{small}
\begin{gather}
    \text{Precision} = \frac{TP}{TP + FP} \label{eq:precision} \\[2ex]
    \text{Recall} = \frac{TP}{TP + FN} \label{eq:recall} \\[2ex]
    \text{F1-score} = \frac{2 \times \text{Recall} \times \text{Precision}}{\text{Recall} + \text{Precision}} \label{eq:f1score} \\[2ex]
    \text{FPR} = \frac{FP}{FP + TN} \label{eq:pf} \\[2ex]
    \text{G-measure} = \frac{2 \times \text{Recall} \times (1 - \text{FPR})}{\text{Recall} + (1 - \text{FPR})} \label{eq:gmeasure}
\end{gather}
\end{small}

All these metrics range from 0 to 1. Among the five performance metrics, \textit{Recall}, \textit{Precision}, \textit{F1-score}, and \textit{G-measure} are preferred to be higher, indicating better performance, while the \textit{FPR} is preferred to be lower.

\begin{table*}[htpb]
\caption{Performance of prompted LLMs, GPT and Gemini (Gem).}
\label{tab:GPTVersusGemini}
\renewcommand{\arraystretch}{1.4}
\centering
\footnotesize
\begin{tabular}{|l|cc|cc|cc|cc|cc|}
\hline
\multirow{2}{*}{\textbf{Datasets}} & \multicolumn{2}{c}{\textbf{Recall}} & \multicolumn{2}{c}{\textbf{Precision}} & \multicolumn{2}{c}{\textbf{F1-score}} & \multicolumn{2}{c}{\textbf{FPR}} & \multicolumn{2}{c|}{\textbf{G-measure}} \\ \cline{2-11}
 & \textbf{GPT} & \textbf{Gem} & \textbf{GPT}  & \textbf{Gem} & \textbf{GPT}  & \textbf{Gem} & \textbf{GPT}  & \textbf{Gem} & \textbf{GPT} & \textbf{Gem} \\ \hline \hline
Chromium & 0.32 & 0.91 & 0.08 & 0.09 & 0.13 & 0.17 & 0.08 & 0.19 & 0.48 &  0.86\\ \hline
Derby & 0.26 & 0.68 & 0.56 & 0.44 & 0.35 & 0.53 & 0.05 & 0.21 & 0.41 &  0.73 \\ \hline
Camel & 0.13 & 0.72 & 0.23 & 0.22 & 0.17 & 0.34 & 0.04 & 0.25 & 0.23 & 0.73 \\ \hline
Ambari & 0.25 & 0.63 & 0.27 & 0.16 & 0.26 & 0.26 & 0.02 & 0.11 & 0.40 & 0.74 \\ \hline
Wicket & 0.17 & 0.78 & 0.18 & 0.19 & 0.18 & 0.31 & 0.04 & 0.16 & 0.29 & 0.81 \\ \hline \hline
\rowcolor{gray!10}
\textbf{Average} & \textbf{0.23} & \textbf{0.74} & \textbf{0.26} & \textbf{0.22} & \textbf{0.22} & \textbf{0.36} & \textbf{0.05} & \textbf{0.18} & \textbf{0.36} & \textbf{0.77} \\ \hline
\end{tabular}
\end{table*}

\section{RESULTS}
\label{sec:EXPERIMENTRESULTS}

We evaluated both prompted proprietary and fine-tuned LLMs for the task of SBR prediction.
In this section, we initially report the performance of the prompted proprietary models and then present and analyze the results of fine-tuned models.

\begin{table*}[htpb]
\caption{Comparison of BERT (B), DistilBERT (DisB), DistilGPT-2 (DisG), and Qwen2.5B (Q).}
\label{tab:CombinedTransformer}
\renewcommand{\arraystretch}{1.4}
\setlength{\tabcolsep}{3.5pt}
\centering
\footnotesize
\resizebox{\textwidth}{!}{%
\begin{tabular}{|l|cccc|cccc|cccc|cccc|cccc|}
\hline
\multirow{2}{*}{\textbf{Datasets}} & \multicolumn{4}{c}{\textbf{Recall}} & \multicolumn{4}{c}{\textbf{Precision}} & \multicolumn{4}{c}{\textbf{F1-score}} & \multicolumn{4}{c}{\textbf{\textbf{FPR}}} & \multicolumn{4}{c|}{\textbf{G-measure}} \\ \cline{2-21}
 & \textbf{B} & \textbf{DisB} & \textbf{DisG} & \textbf{Q} & \textbf{B} & \textbf{DisB} & \textbf{DisG} & \textbf{Q} & \textbf{B} & \textbf{DisB} & \textbf{DisG} & \textbf{Q} & \textbf{B} & \textbf{DisB} & \textbf{DisG} & \textbf{Q} & \textbf{B} & \textbf{DisB} & \textbf{DisG} & \textbf{Q} \\ \hline \hline
Chromium & 0.70 & 0.74 & 0.79 & 0.77 & 0.90 & 0.83 & 0.84 & 0.87 & 0.78 & 0.79 & 0.82 & 0.82 & 0.00 & 0.00 & 0.00 & 0.00 & 0.83 & 0.85 & 0.88 & 0.87 \\ \hline
Derby & 0.40 & 0.39 & 0.10 & 0.28 & 0.68 & 0.78 & 0.12 & 0.45 & 0.51 & 0.52 & 0.11 & 0.34 & 0.04 & 0.03 & 0.18 & 0.08 & 0.57 & 0.56 & 0.18 & 0.43 \\ \hline
Camel & 0.20 & 0.20 & 0.02 & 0.15 & 0.60 & 0.53 & 0.04 & 0.35 & 0.30 & 0.29 & 0.03 & 0.21 & 0.01 & 0.02 & 0.05 & 0.03 & 0.33 & 0.33 & 0.04 & 0.26 \\ \hline
Ambari & 0.25 & 0.19 & 0.06 & 0.13 & 0.31 & 0.60 & 0.04 & 0.08 & 0.28 & 0.29 & 0.05 & 0.10 & 0.02 & 0.00 & 0.05 & 0.05 & 0.40 & 0.32 & 0.12 & 0.22 \\ \hline
Wicket & 0.13 & 0.30 & 0.02 & 0.17 & 0.43 & 1.00 & 0.10 & 0.80 & 0.20 & 0.47 & 0.04 & 0.29 & 0.01 & 0.00 & 0.02 & 0.00 & 0.23 & 0.47 & 0.04 & 0.30 \\ \hline \hline
\rowcolor{gray!10}
\textbf{Average} & \textbf{0.34} & \textbf{0.36} & \textbf{0.20} & \textbf{0.30} & \textbf{0.58} & \textbf{0.75} & \textbf{0.23} & \textbf{0.51} & \textbf{0.41} & \textbf{0.47} & \textbf{0.21} & \textbf{0.35} & \textbf{0.02} & \textbf{0.01} & \textbf{0.06} & \textbf{0.03} & \textbf{0.47} & \textbf{0.51} & \textbf{0.25} & \textbf{0.42} \\ \hline
\end{tabular}}
\end{table*}

\begin{tcolorbox}[RQsboxstyle]
    \textbf{RQ\textsubscript{1}:} How effective are proprietary LLMs at predicting SBRs?
\end{tcolorbox}



Table~\ref{tab:GPTVersusGemini} summarizes the performance of GPT and Gemini on SBR prediction across the test splits of all five datasets. Reported metrics are macro-averaged over the datasets. Gemini consistently outperforms GPT, achieving a higher average F1-score and G-measure.

Gemini achieves substantially higher recall than GPT, with an average recall of 0.74 vs. 0.23, detecting over three times as many SBRs on average.
The recall advantage is then reflected in the G-measure and F1-score, where Gemini achieves 0.77 and 0.36, respectively, compared to 0.36 and 0.22 for GPT.
GPT is more Conservative as it misses the vast majority of SBRs but rarely raises a false alarm with a FPR of only ~0.05 (5\%).
Gemini, in contrast, applies a less conservative methodology, resulting in a higher FPR of 0.18 (18\%) and a higher likelihood of misidentifying NSBR instances as SBR.

\begin{tcolorbox}[yellowboxstyle]
Gemini detects over three times more SBRs than GPT, leading to a higher F1-score and G-measure across datasets. However, this gain in sensitivity is accompanied by increased FPR.
\end{tcolorbox}

In each dataset, a consistent trend emerges, with Gemini demonstrating substantially higher recall and G-measure values. In terms of F1-score, Gemini outperforms GPT in four out of five datasets and matches GPT on Ambari. Precision results are mixed, with GPT achieving higher values on Derby, Camel, and Ambari, while Gemini slightly surpasses GPT on Chromium and Wicket.

For instance, on the large-scale Chromium dataset, Gemini achieved a notably higher recall of 0.91 compared to GPT at 0.32, albeit with a higher FPR (0.19 vs. 0.08). This recall advantage persisted across the smaller datasets as well; for example, Gemini attained recall scores of 0.72 on Camel and 0.78 on Wicket, substantially exceeding GPT’s corresponding values of 0.13 and 0.17. Although GPT consistently exhibited higher precision, as seen in Derby (0.56 vs. 0.44) and Ambari (0.27 vs. 0.16), it showed a limited recall of 0.32 or lower in all the other cases. As a result, Gemini consistently produced superior G-measure across all datasets, often by a considerable margin. Notably, it reached 0.86 on Chromium and 0.81 on Wicket, indicating that despite generating more false positives, it delivers broader and more effective coverage for security auditing tasks.

\begin{tcolorbox}[yellowboxstyle]
In each dataset, Gemini proves to be a more effective tool for SBR prediction; its superior G-measures (reaching 0.86 on Chromium) indicate better performance when identifying SBR is prioritized over minimizing false alarms.
\end{tcolorbox}

\begin{tcolorbox}[RQsboxstyle]
    \textbf{RQ\textsubscript{2}:} How effective are fine-tuned LLMs at predicting SBRs?
\end{tcolorbox}

We evaluated four fine-tuned models, including BERT, DistilBERT, DistilGPT-2, and Qwen2.5B for the task of SBR prediction. The results are summarized in Table~\ref{tab:CombinedTransformer}.


DistilBERT consistently outperforms all other models across most evaluation metrics. It achieves the highest average recall (0.36), F1-score (0.47), and G-measure (0.51), indicating better performance in identifying SBRs while maintaining a balanced trade-off between precision and recall. It also achieves the highest average precision of 0.75.
BERT also demonstrates competitive performance, with an average recall of 0.34 and an F1-score of 0.41, and achieves the second-highest average precision (0.58) among all models. 
These results suggest that encoder-based LLM architectures are more effective at capturing the features necessary for SBR prediction.

\begin{tcolorbox}[yellowboxstyle]
Among the fine-tuned LLMs, DistilBERT achieves the best average performance across all key metrics, including Recall, Precision, F1-score, and G-measure in predicting SBRs.
\end{tcolorbox}

In contrast, DistilGPT exhibits substantially lower average recall (0.20), F1-score (0.21), and G-measure (0.25).
While Qwen outperforms DistilGPT, it still lags behind encoder-based models in recall (0.30) and F1-score (0.35), though it achieves a relatively high average precision (0.51). This suggests that these models tend to be more conservative, favoring precision at the expense of recall.

Across individual datasets, the four models exhibit distinct performance evaluations.
All models show strong performance on the Chromium dataset; however, DistilGPT-2 and Qwen achieve the highest recall (0.79 and 0.77, respectively) and F1-scores (both at 0.82), outperforming the others. 
Derby is the second most favorable dataset, with all models achieving substantially higher recall and F1-scores compared to their performance on the other datasets.
Interestingly, while Chromium has the highest total number of bug reports, the ratio of SBRs to NSBRs is relatively small (1.92\%). Conversely, Derby, though it has fewer overall bug reports than Chromium, has the highest relative frequency of SBRs compared to NSBRs (17.9\%).
However, on more challenging datasets such as Camel and Ambari, all models experience a marked drop in recall and F1-score, highlighting the inherent difficulty of SBR prediction in these contexts.

\begin{tcolorbox}[yellowboxstyle]
Individual datasets exhibit substantial variability in model effectiveness; all models achieve their best scores on Chromium and Derby, yet encounter pronounced difficulties with Camel, Ambari, and Wicket.
\end{tcolorbox}

\section{Discussion}
\label{sec:Discussion}

In this section, we discuss the broader implications of prompting and fine-tuning LLMs for SBR prediction, drawing insights from our experimental results.

\subsection{Zero-shot and Few-shot Prompting.}
The classification performance of GPT and Gemini models is known to be affected by prompt design~\cite{11052790}.
As such, we also investigated the impact of different prompt engineering techniques, including zero-shot and few-shot techniques.
In the few-shot setting, examples in the prompt were randomly sampled from the training set of each dataset, avoiding any use of test sets to prevent data leakage and ensure a reliable estimate of model performance.
Our results showed no significant performance difference between few-shot and zero-shot settings, as confirmed by paired t-test analysis. In particular, we did not observe a performance improvement when examples were provided in the prompt, compared to the setting in which the prompt only contained the definition of labels used for prediction.

For instance, on the Wicket dataset, Gemini achieved an F1-score of 0.31 in the zero-shot setting, which slightly decreased to 0.29 with few-shot prompting. Similarly, GPT achieved an F1-score of 0.22 in the zero-shot setting, dropping to 0.18 in the few-shot setting.
A similar trend was observed on the Camel dataset. For Gemini, the F1-score was 0.34 in the zero-shot setting and decreased slightly to 0.29 in the few-shot setting. In contrast, GPT’s F1-score improved from 0.17 (zero-shot) to 0.22 (few-shot). However, these variations were not statistically significant, as confirmed by a paired t-test ($p > 0.05$).


\subsection{Model Agreement in Proprietary LLMs.}
To assess the level of agreement between the two prompted models, GPT and Gemini, we computed a consensus score for each sample across all datasets. This score ranges from 0 (neither model correctly predicts the label) to 2 (both models correctly predict the label).
Figure~\ref{fig:consistencyScore} illustrates the distribution of these scores across the datasets. 


\definecolor{pastelgreen}{HTML}{A3C4BC} 
\definecolor{pastelorange}{HTML}{E7D0A6} 
\definecolor{pastelred}{HTML}{BF8A88}   

\begin{figure}[ht]
    \centering
    \begin{tikzpicture}
        \begin{axis}[
            ybar stacked,
            bar width=18pt,
            width=\columnwidth,
            height=6.3cm,
            symbolic x coords={Chromium, Derby, Ambari, Camel, Wicket},
            xtick=data,
            xticklabel style={font=\footnotesize},
            enlarge x limits=0.15,
            ymin=0, ymax=115, 
            ylabel={Percentage (\%)},
            legend style={
                at={(0.5,-0.15)},
                anchor=north,
                legend columns=-1,
                draw=none,
                fill=none,
                font=\footnotesize
            },
            nodes near coords={%
                \pgfmathprintnumber[fixed, precision=1]{\pgfplotspointmeta}\%%
            },
            every node near coord/.append style={
                font=\tiny, 
                color=black!80, 
                yshift=-1pt 
            },
            axis line style={draw=none}, 
            tick style={draw=none},      
            ymajorgrids=true,            
            grid style={dashed, gray!30},
            cycle list={
                {fill=pastelgreen, draw=pastelgreen!50!black, thick}, 
                {fill=pastelorange, draw=pastelorange!50!black, thick}, 
                {fill=pastelred, draw=pastelred!50!black, thick}
            }
        ]
            \addplot coordinates {(Chromium, 71.08) (Derby, 68.60) (Ambari, 87.20) (Camel, 68.20) (Wicket, 80.60)};
            
            \addplot coordinates {(Chromium, 25.67) (Derby, 21.40) (Ambari, 9.60) (Camel, 26.20) (Wicket, 15.20)};
            
            \addplot+[
                nodes near coords style={
                    yshift=-4.3pt,  
                    anchor=south, 
                    color=black   
                }
            ] coordinates {(Chromium, 3.25) (Derby, 10.00) (Ambari, 3.20) (Camel, 5.60) (Wicket, 4.20)};

            \legend{Both Correct (2), One Correct (1), None Correct (0)}
        \end{axis}
    \end{tikzpicture}
        \caption{Prediction consistency scores between GPT and Gemini across five datasets.}
    \label{fig:consistencyScore}
\end{figure}
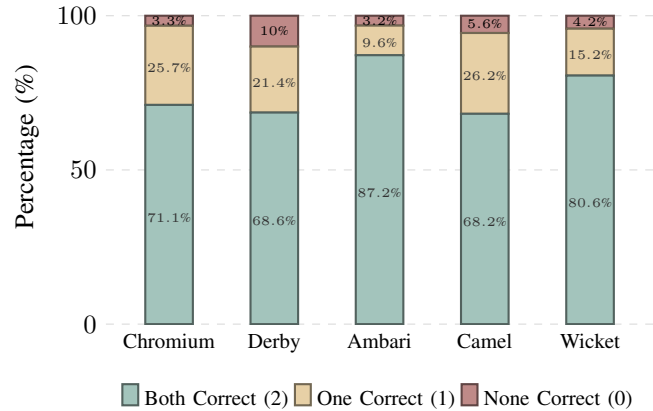


The results demonstrate that across all datasets, Score-2 is the dominant category, ranging from roughly 68\% to 87\%, with both models in consensus in identifying the correct labels.
Ambari is the easiest dataset, with an agreement rate of 87.2\%, followed by similarly strong results on Wicket. 
Score-1 Cases, where only one model is correct, account for approximately 10\% to 26\% of cases and are notable in Camel (26.2\%) and Chromium (25.7\%).
Score-0 cases are rare across all datasets (typically below 6\%) except Derby, which reaches approximately 10\%, indicating that neither model could predict the correct labels. 
In this dataset, 50 cases (10\%) were misclassified, comprising 31 FNs and 19 FPs. 

Overall, although cases where both models are incorrect are rare, it is not uncommon for only one model to be correct. This suggests that combining model predictions could further improve results and represents a promising direction for future research.

%

\subsection{Hyperparameter Optimization for Fine-Tuned Models.}
During hyperparameter tuning across different models, we evaluated model sensitivity to maximum sequence length and learning rate, revealing distinct performance trends across the datasets. We have reported the best results, but further optimization may still yield improvements under certain conditions. 
The optimal sequence length was highly dataset-dependent. For instance, for Chromium, Camel, and Derby, a maximum length of 256 tokens using Qwen consistently yielded the highest performance. On the Chromium dataset, increasing the length from 256 to 512 did not improve the metrics (F1-score dropped from 0.91 to 0.90) but significantly increased training time from 6.73 hours to 7.38 hours. Similarly, on Camel, the 256-length setting achieved an F1-score of 0.57, far surpassing both the 128-length (F1-score 0.47) and 512-length (F1-score 0.53) configurations.

\subsection{Prompted vs. Fine-tuned Models.}
Although Gemini achieved a high G-measure and recall, at 0.77 and 0.74, respectively, its high FPR resulted in critically low precision (e.g., 0.09 on Chromium). On the other hand, DistilBERT outperformed all other fine-tuned models with an F1-score of 0.47 and a G-measure of 0.51. 
To provide a more comprehensive comparison, we also examine execution time and cost, assessing the efficiency and resource requirements of each approach.

\emph{Execution Time.} In terms of execution time, GPT demonstrated a clear advantage over Gemini across all evaluated datasets. On smaller datasets such as Derby and Ambari, GPT processed each row in approximately 1.7 seconds, completing the full dataset in under 15 minutes, whereas Gemini required 3.6–4.3 seconds per row, with total execution times exceeding 29 minutes. This disparity became more pronounced on larger datasets like Chromium, where GPT completed processing in approximately 19.5 hours, while Gemini required over 50 hours. Overall, GPT consistently achieved faster per-row processing and substantially lower total execution times, indicating higher computational efficiency, which is particularly beneficial in large-scale SBR prediction tasks. 

In contrast, the fine-tuned architectures require an initial computational investment for training. For instance, training the models on the largest dataset, Chromium, required between 2.86 and 7.38 hours, depending on the maximum sequence length. However, this upfront cost is heavily compensated by superior inference speeds after deployment. Once trained, the fine-tuned models processed the entire Chromium dataset (approx. 20,900 rows) in just 10 to 47 minutes, compared to the 19.5 hours required by GPT and 50 hours by Gemini. This translates to an average inference latency of roughly 0.03–0.13 seconds per row, representing a speedup of over $10\times$ to $50\times$ compared to the 1.7–4.3 seconds per row seen with the large-scale LLMs.

\emph{Cost Analysis.} GPT exhibited a substantial advantage in terms of financial cost in comparison to Gemini. The total expenditure for running the experiments with GPT was approximately \$25, whereas the same workload processed by Gemini incurred a cost of \$71. This nearly threefold increase in costs makes Gemini significantly less economical for large-scale deployments. While Gemini offers superior recall, GPT provides a far more budget-friendly solution for organizations prioritizing cost minimization.

In terms of costs, training fine-tuned small LLMs is possible on a local machine, requiring only moderate computational resources and completing within a reasonable time. For instance, training the BERT model on Google Colab (using an NVIDIA T4 GPU) on the Chromium dataset required approximately seven hours per train–test split, whereas the other datasets finished within fewer hours.

\subsection{Ethical and Operational Implications.} 
Applying LLMs to security-sensitive bugs introduces ethical and operational considerations that extend beyond model accuracy. From an ethical perspective, transmitting potential security bug report descriptions to third-party API providers (e.g., OpenAI, Google) raises issues of data privacy and sovereignty. The description, essential for effective classification, may contain proprietary logic or architectural details. By contrast, local models such as BERT, which we trained once and reused without external data transfer, offer a privacy-preserving alternative that keeps all sensitive data within an organization’s internal infrastructure.


\section{THREATS TO VALIDITY}
\label{sec:ThreatsToValidity}

\emph{Internal validity.}
Hyperparameter tuning across different models revealed distinct performance trends across the datasets. Although we have reported the best results, these findings may not generalize, as alternative hyperparameter tuning approaches and settings could produce different outcomes.
Using the same parameter settings (temperature = 0.2, top p = 0.1) across prompt-based models may further affect internal validity, as observed differences could partially reflect sensitivity to these parameter choices rather than inherent differences between models \cite{openai2023cheatsheet}.
Further experimentation may lead to improved results. While our study may not achieve optimal SBR prediction, it highlights the benefits of prompt-based and fine-tuned models. However, the question of how larger LLMs build an effective model remains open for investigation.

\emph{External validity.} Our analysis is limited to the five datasets Chromium, Derby, Ambari, Camel, and Wicket, which we did not update to the latest version to ensure a fair comparison with state-of-the-art studies. This may restrict the applicability of our results to current trends.
To generalize our findings, replication across multiple projects from different domains with labeled bug reports is required. Furthermore, while we focused on six models for fine-tuning and prompting LLM models, other variants could potentially offer improved performance.
The dataset-specific nature of optimal parameters may also limit the generalizability of our findings to other datasets or future model versions.

\emph{Conclusion validity.}  Threats can impact the accuracy of our conclusions. Comparing learning system performance is challenging due to the variety of available metrics.

\section{Conclusion}
\label{sec:Conclusion}

We benchmark the capabilities of prompted versus fine-tuned Large Language Models (LLMs) for Security Bug Report (SBR) prediction on five publicly available datasets.
We find that Gemini, while achieving the highest G-measure of 77\%, suffers from high false positive rates (FPRs), excessive inference latency, and operational costs that are nearly three times higher than those incurred by GPT. On the other hand, GPT offers better computational and financial efficiency, but its emphasis on precision leads to many missed SBRs and a low G-measure of 36\%.
In contrast, fine-tuned models yield a lower average G-measure, with the best-performing model, DistilBERT, reaching 51\% but substantially
higher precision of 75\% at the cost of reduced recall of 36\%. However, these models, once trained, offer 10 to 50 times faster inference than proprietary LLMs.
These findings warrant more in-depth investigations to understand the reasons for the limited performance of models for SBR prediction.

\section*{Data Availability}

The dataset and training scripts are publicly available: \url{https://zenodo.org/records/18290275}

\balance
\bibliographystyle{IEEEtran}
\bibliography{references}

\end{document}